# White Paper

towards a fuller understanding of

# ICY SATELLITE SEAFLOORS, INTERIORS, AND HABITABILITY

**Paul Byrne** NC STATE UNIVERSITY | paul.byrne@ncsu.edu

**Andrew Dombard** UNIVERSITY OF ILLINOIS AT CHICAGO

**Catherine Elder** JET PROPULSION LABORATORY, CALTECH

**Steven Hauck, II** CASE WESTERN RESERVE UNIVERSITY

**Mohit Melwani Daswani** JET PROPULSION LABORATORY, CALTECH

**Paul Regensburger** UNIVERSITY OF OREGON

**Steven Vance** JET PROPULSION LABORATORY, CALTECH

**ENDORSERS** CAITLIN AHRENS | MICHELE BANNISTER | ALEXIS BOUQUET | TRISTAN CARO | ELLEN CZAPLINSKI | ADEENE DENTON | MARSHALL EUBANKS | MARK FOX-POWELL | CESARE GRAVA | JAMES HABER | JENNIFER HANLEY | HAMISH HAY | BAPTISTE JOURNAUX | STEPHEN KANE | JAMES KEANE | SCOTT KING | CHRISTIAN KLIMCZAK | SIDDHARTH KRISHNAMOORTHY | ALICE LUCCHETTI | PATRICK MCGOVERN | ALYSSA MILLS | ERICA NATHAN | JULIE NEKOLA NOVÁKOVÁ | JESSICA NOVIELLO | JOSEPH O'ROURKE | MAURIZIO PAJOLA | MARK PANNING | RUTU PAREKH | DIVYA PERSAUD | EDGARD RIVERA-VALENTÍN | PAUL SCHENK | ASHLEY SCHOENFELD | LAUREN SCHURMEIER | HEATHER SMITH | KRISTA SODERLUND | SIMON STÄHLER | MARSHALL STYCZINSKI | DAVID TOVAR | KEVIN TRINH | ANNE VERBISCER | MAHEENUZ ZAMAN

**Key Findings**
- **Icy satellites represent compelling exploration target as potential habitable worlds, but the properties of their rocky interiors must be better characterized and more fully considered**
- **Funding fundamental research programs and thematic workshops that promote ocean world interdisciplinarity is key, with scientists from multiple backgrounds working to develop holistic interior–seafloor–ocean–ice shell models**
- **Future missions to icy satellites should explicitly include objectives to characterize rock–water/high-pressure ice interfaces via established geophysical methods such as gravity science**
- **The payoff for this interdisciplinarity will be considerable, offering an enhanced insight into the potential habitability of these worlds, and of rock–water environments more broadly**

## 1. Introduction

The icy satellites of the Solar System's giant planets are key astrobiological targets. Featuring known or suspected subsurface oceans (Pappalardo *et al.*, 1999; Iess *et al.*, 2012; Thomas *et al.*, 2016), likely situated atop rocky interiors (e.g., Anderson *et al.*, 1998), these moons may therefore harbor chemoautotrophic habitable environments at their seafloors—where liquid water, biochemically vital elements (such as C, H, O, S, N, and P), and energy sources may be present together (Hoehler *et al.*, 2007). Indeed, the detection of silicate grains in Enceladus' plumes provides strong evidence for rock–water interactions within that diminutive moon (Hsu *et al.*, 2015), and invoking a wet, porous, permeable, and unconsolidated rocky interior accounts for the sustained presence of a subsurface ocean there (Choblet *et al.*, 2017). Given that life on Earth may have originated at sites of serpentinization where seawater and basalt interacted (e.g., Martin *et al.*, 2008), it stands to reason that we look to these ocean worlds to better understand how life might arise generally.

### 1.1. Astrobiological Targets

This focus on icy satellite astrobiology underpins a major part of NASA's spacecraft exploration strategy. For example, the upcoming Europa Clipper mission will characterize the potential habitability of that satellite by measuring physical and chemical properties of the ice shell and subsurface ocean (e.g., Howell and Pappalardo, 2020). The Discovery-class Trident mission concept (Mitchell *et al.*, 2019) would, if selected, verify the existence of an extant liquid water ocean beneath Triton's icy surface. And the Outer Planets Assessment Group (OPAG) report *Scientific Goals for Exploration of the Outer Solar System* notes "Determining the presence and natures of sub-surface oceans, especially whether liquid water is in direct contact with rock interiors as is suspected at Enceladus and Europa, is crucial to understanding the evolution and potential habitability of these bodies and how materials are processed within them" (version dated 2019-08-28).

### 1.2. The Inaccessible Interior

Alas, these deep interiors are not directly accessible, and so considerable inferences must be made about conditions there. The rocky seafloors within these satellites range from tens (e.g., Enceladus) to several hundreds (e.g., Ganymede) of kilometers below their icy surfaces (*Figure 1*)—far below the deepest oceans on Earth. Further, beneath those cold carapaces are unique rocky worlds in their own right, perhaps considerably different to those of the inner Solar System with which we are much more familiar. Yet these deep rocky layers are responsible, at a minimum, for a source of heat to the base of these oceans from radiogenic decay and/or tidal dissipation. Characterizing the properties of and processes operating within these deep interiors—including their composition, rheology, and melting behavior, as well as the propensity for, and styles of, mantle convection, magmatism, and tectonics—is essential for understanding the evolution of these bodies and their potential physical and chemical interactions with the oceans that lie above them. **The inaccessibility of these deep interiors and the sparse geophysical data available for them severely limits our grasp of ocean world habitability.**

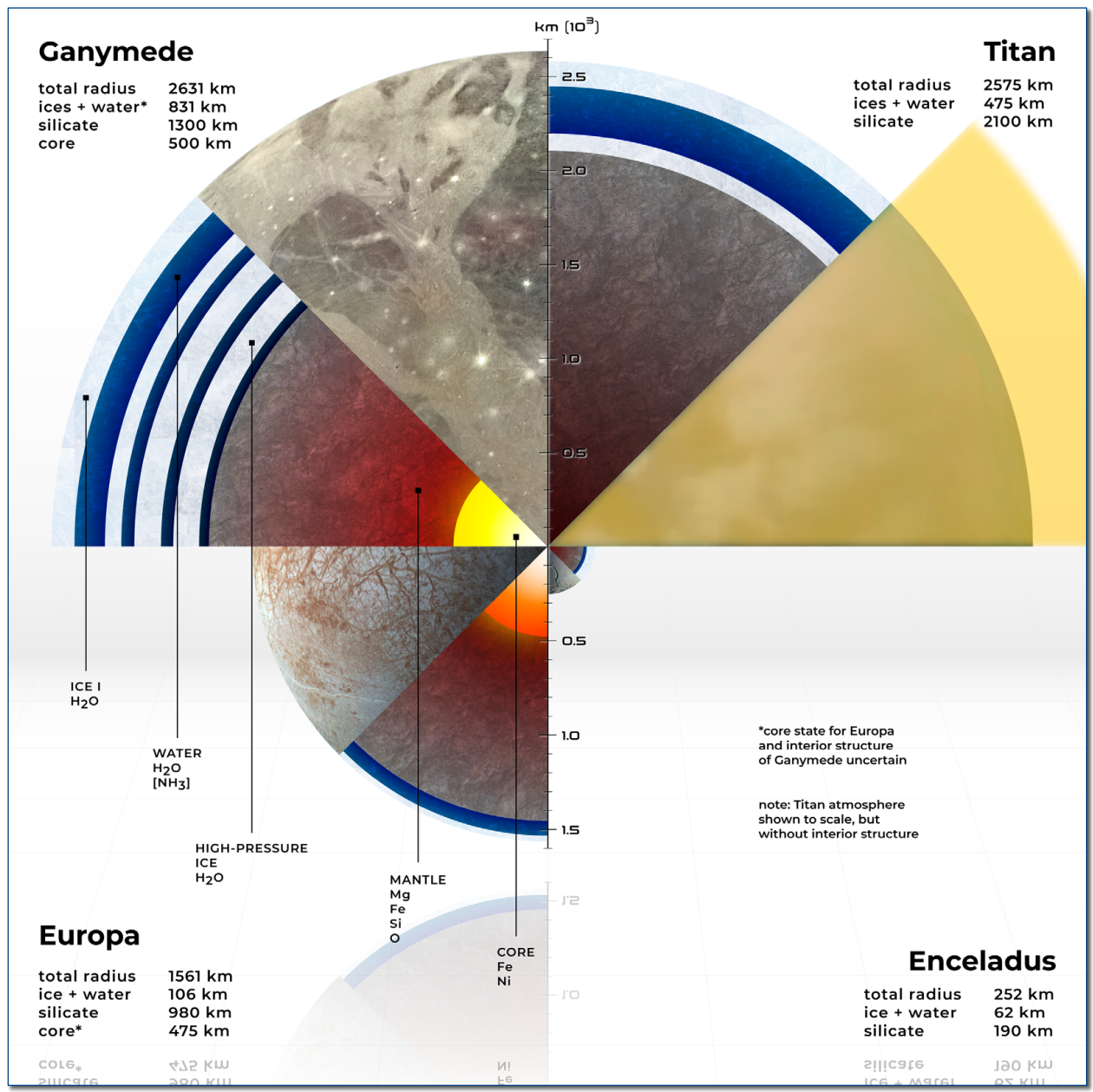

*Figure 1. The interior structures of select icy satellites, to scale. Clockwise from top left: Ganymede, Titan, Enceladus, and Europa. Interior structures are taken from* Vance et al. *(2017). The state of Europa's core (molten or solid) is uncertain; the interior structure of water and ice layers within Ganymede is representative only. Note that the Titan atmosphere is shown for completeness, but without its own interior structure (e.g., haze layers).*

## 2. Alien Worlds Under Alien Seas
### 2.1. Under Pressure

One approach has been to anticipate that geophysical processes at the seafloors of these worlds are similar in nature to those that support chemoautotrophic environments on Earth. Although not an *explicit* assumption, perhaps, illustrations of these rock–water interfaces in NASA press releases for instance often feature hydrothermal vent systems at mid-ocean ridges on Earth (***Figure 2***). But on Earth, these systems are augmented by the presence of fractures (Pirajno, 2009; Magee *et al*., 2015), and the ultimate engine of such activity is plate tectonics.

Although rock–water interactions are common in ocean settings on Earth, it is not clear that similar physical and geological conditions are readily applicable to every ocean world. For example, in Enceladus' rocky interior, the percolation of water through porous silicate rock believed to underpin the chemistry of the plumes emitting from that moon would facilitate rock–water interactions within interstitial pore spaces. However, such porous flow may not exist, or even be possible, in larger icy worlds. Pressures at the interfaces between water (or water ice) and the underlying silicates may close pore spaces and choke off permeability, thereby limiting water–rock interactions at the direct interface and thus greatly inhibiting the surface area available for chemical reactions (Byrne *et al*., 2018).

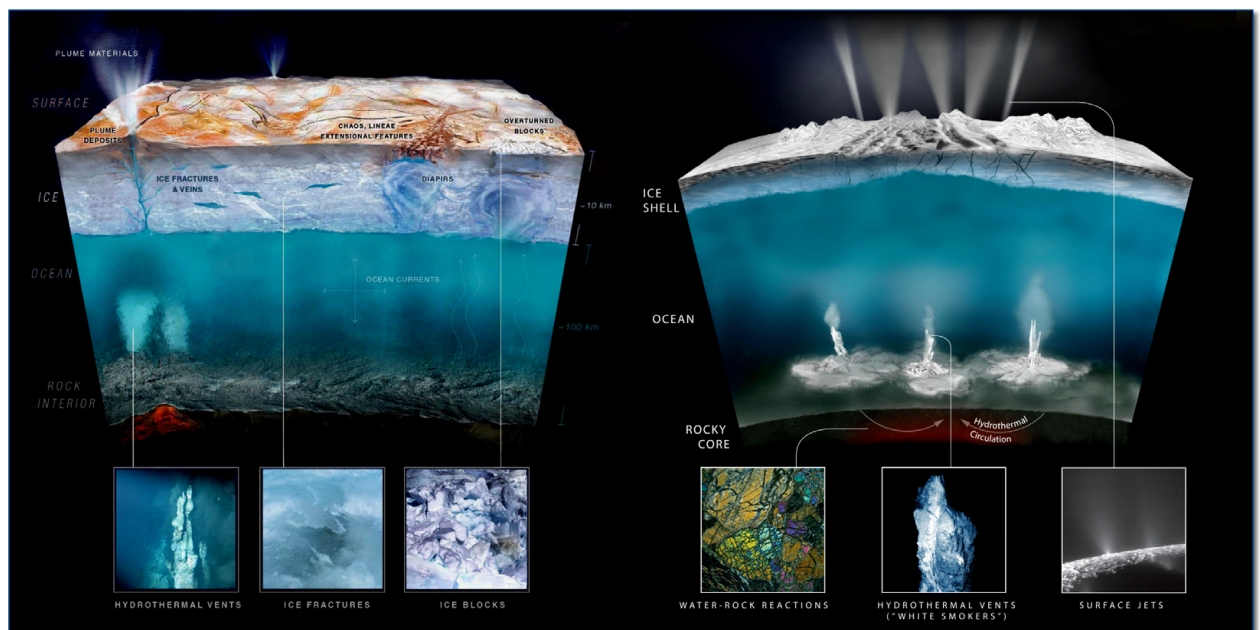

***Figure 2.*** *Artists' impressions of the interiors of Europa (left) and Enceladus (right). Hydrothermal vents feature on both the Europan and Enceladan seafloors (labeled "white smokers" for the latter). Although there appears to be some form of hydrothermal circulation at the Enceladan seafloor, the style of such behavior is unknown—and there is as yet* no *evidence for hydrothermal vents within Europa. Image credits: NASA/JPL-Caltech.*

To illustrate this point, a 100 km-deep ocean (or mix of ice and liquid water layers) within an icy world for which the surface gravitational acceleration is ~1–2 m/s$^2$—conditions eminently applicable to Europa, Titan, and Ganymede—results in pressures of order 100 MPa at the seafloor. When compared with typical rock failure strengths (~10 MPa), it naturally follows that fracturing will be inhibited. The reality is probably more complicated, with any pores at the seafloor likely water filled and thus the rock there at hydrostatic pressure and so relatively weak (Klimczak *et al*., 2019). Even so, within a few hundred meters below the seafloor, as conditions tend towards lithostatic pressure, the stresses needed to drive frictional sliding may quickly exceed any available driving mechanism.

Even our current estimates of ocean compositions rely on the assumed but unknown composition of the rocky interior (typically based on ~4.5 Gyr-old meteorites) and efficient water–rock interaction that has thermodynamically equilibrated the water layer with the underlying rock (e.g., Zolotov and Kargel, 2009). However, as we discuss, water–rock interaction may be severely limited in these environments, precluding the nutrient and redox cycling necessary to maintain metabolic processes in a light-starved ocean (Gaidos et al., 1999). **As a result, commonly held expectations based on direct** **analogy to the bottoms of Earth's oceans for the physical and chemical processes within these larger icy worlds are, at best, equivocal.**

## *2.2. The Prospect for Geological Activity?*

It is possible, however, that even with high pressure limiting the fracturing and frictional sliding of rock, other geological processes may be at work within icy satellites. Magma ascent is typically thought of as taking place through fracture networks acting as conduits through otherwise intact and impermeable rock. But the migration of magma at depth from the zone of partial melting (say, at the top of a mantle plume) to the base of the brittle lithosphere is in fact accommodated by corrosive etching through the upper mantle (e.g., Spiegelman *et al.*, 2001; Beck *et al.*, 2006). High lithostatic pressures clearly do not preclude such processes on Earth, and so perhaps silicate melt within the rocky interiors of ocean worlds could corrode all the way to the rock–water interface, erupting at the seafloor.

Yet predicting whether melting of these rocky interiors even took place in the first place requires a leap in our understanding of where heat is actually generated in icy satellites, which is intimately tied to the interior's rheological properties (e.g., Peale, 2003). The rheological properties, in turn, directly control how and where tidal energy is dissipated in Europa (e.g., Hussmann and Spohn, 2004; Moore and Hussmann, 2009) and Enceladus (e.g., Meyer and Wisdom, 2007; Roberts, 2015), and possibly also Ganymede and Titan. Establishing if seafloor volcanism has operated on ocean world seafloors bears implications for the geological and geophysical condition there (e.g., in terms of topography), as well as for mineral and nutrient exchange between the interior and the ocean.

Those dark, cold seafloors might also still show evidence of impact bombardment, especially if the water/ice overburden is sufficiently thin. The transient crater of an Orientale-basin scale impact (i.e., forming a ~900 km-diameter impact feature) temporarily excavates to depths of hundreds of kilometers (Potter *et al.*, 2013), enough to reach through Europa's ocean (e.g., Vance *et al.*, 2017) and resulting in a cratered landscape at least at large scales (cf. Dombard and Sessa, 2019). Yet even smaller impactors, likely to have punctured Europan ice shell at intervals of tens to hundreds of millions of years (Cox and Bauer, 2015), could have deposited material onto the seafloor. The presence of such impact-generated topography, in turn, suggests that mechanical weathering, aided by water–rock interactions and even by mass wasting (e.g., submarine landslides), may shape these ocean floors.

Taken together, the silicate cores of icy worlds may not be geologically dead, but any activity there could be very episodic, with short-lived volcanic, tectonic, or mass-wasting phenomena separated by extended periods of quiescence. Such a scenario of punctuated geological activity has major implications for the supply of nutrients to icy satellite seafloors, and the prospect for these environments *remaining* habitable over extended periods of time (i.e., tens to hundreds of millions of years). It is clear, therefore, that more careful consideration of possible processes at work on icy satellite seafloors is clearly needed, and that **at present we have an incomplete basis for fully evaluating the habitability of these worlds.**

## 3. The Next Steps Forward

It is critical that the planetary science community explicitly considers the rocky interiors and rock–water interfaces in current and future studies of ocean worlds. To do so successfully requires true interdisciplinarity, whereby scientists with backgrounds encompassing expertise in fracture mechanics, marine geophysics, volcanism, geochemistry and ocean chemistry, and biology (among others) are brought together. **We thus encourage the planetary science community to take a holistic view of these worlds by closely integrating multiple disciplines, especially by including workers from the Earth marine geoscience community.**

To do so, expanded NASA support for interdisciplinary meetings and funding opportunities—such as, for instance, the Habitable Worlds and Exoplanet Research programs—will allow for increased interactions between scientists who might not otherwise have the opportunity or ability to collaborate. Such programs are suitable vehicles for supporting analytical, numerical, and laboratory investigations of key outstanding science questions for ocean world interiors, including but not limited to:

- the prospect for, and type(s) of, geological activity within and at the surfaces of the rocky portions of icy satellites, and whether that activity has any counterpart on Earth or elsewhere in the Solar System;
- the nature and rates of chemical reactions at silicate–high-pressure ice boundaries, such as at the seafloor of Titan and possibly Ganymede; and
- the duration of any redox reactions at icy satellite seafloors, and thus whether chemoautotrophic environments in these settings are sustainable over geological timespans.

Similarly, interdisciplinary conference sessions and thematic workshops will further help foster existing and new collaborations between nominally disparate fields within the geosciences, especially if held as part of larger meetings such as the annual American Geophysical Union Fall Meeting or the European Geosciences Union General Assembly. **NASA-supported efforts that focus at least in part on the rock–water environment would be a major step in building towards a fuller understanding of the habitability of ocean worlds.**

Another crucial component of advancing our understanding of the habitability of ocean worlds is to consider their deep interiors as the rocky planetary bodies they are. Indeed, Ganymede's silicate interior and Titan's rock-rich layers are both as large the Moon, an impact-bombarded body that experienced considerable volcanic and tectonic activity during its first billion years or so (Solomon and Head, 1980; Morota *et al*., 2011). Understanding both the geological histories of the surfaces of these rocky bodies (e.g., volcanism, tectonics, impact cratering, mass wasting, weathering by tidal currents, etc.) and the thermochemical evolution of their interiors is essential. **We therefore advocate for the Decadal Survey to emphasize fundamental research into the operation of planetary interiors with the conditions and properties of the deep interiors of ocean worlds**—building on, but expanding beyond, the better-understood parameter space of the inner Solar System worlds.

Additionally, planned and future spacecraft missions, even if focused primarily on the surfaces of icy satellites, should treat the rock–water/high-pressure ice interface as a fundamental science objective. For instance, although Europa Clipper will mainly assess the habitability of Europa via detailed examination of its ice shell to understand how chemical reactants at the surface might mix into the interior (Howell and Pappalardo, 2020), measurements of anomalous accelerations of the spacecraft could reveal local gravity anomalies at the rock–water interface, the magnitude of which may offer insight into the heat flow from the silicate core (Dombard and Sessa, 2019). The finding of locally elevated heat flows, or even evidence for topography, on the Europan seafloor would dramatically enhance our view of that environment and its geological properties. Geophysical measurements conducted on the surface of the Enceladan ice shell would similarly return critical insights into the interior structure of that moon (Vance *et al*., 2020). **Expanding the goals of surface-focused missions to an icy satellite(s) to include the ocean floor would readily lead to a fuller assessment of ocean world habitability than is possible with observations of the icy shell alone.**

Finally, and as we discuss here, fully characterizing icy satellite habitability requires drawing on perspectives spanning the gamut of planetary science, geoscience, and beyond. Studies of scientific teams have repeatedly demonstrated the importance of an integrated approach, whereby team members with diverse expertise develop synergies between their specialties and resources that result in an end product greater than the sum of its parts (Balakrishan *et al*., 2011). Sociological studies demonstrate that groups fostering strong connections across discipline boundaries are more innovative (Powell et al., 1996; Burt, 2004; de Vaan *et al.* 2015), leading to higher-impact outcomes that endure (Curral *et al.*, 2001; de Vaan *et al.*, 2015). **We strongly encourage the Decadal Survey to consider the state of the profession and the issues of equity, diversity, inclusion, and accessibility—not as separable issues, but as critical steps on the pathway to understanding ocean worlds specifically and the entire Solar System more generally.** Background information on the lack of diversity in our community and specific, actionable, and practical recommendations can be found in white papers by Rivera-Valentín *et al.* (2020), Milazzo *et al.* (2020), Rathbun *et al.* (2020), and Strauss *et al.* (2020).